\documentclass[aps,amssymb,prl,twocolumn]{revtex4}
\usepackage{graphicx}

\begin{document}
\title{Far-infrared electrodynamics of superconducting Nb:\\
comparison of theory and experiment}
\author{A. E. Karakozov \footnote {Electronic address:
karakozov@mtu-net.ru}} \affiliation{L. F. Vereschagin Institute of
High Pressure Physics, Russian Academy of Sciences, 142190
Troitsk, Moscow region, Russia}
\author{E. G. Maksimov \footnote {Electronic address:
maksimov@lpi.ru}} \affiliation{P. N. Lebedev Physical Institute,
Russian Academy of Sciences, 119991 Moscow, Russia}
\author{A. V. Pronin \footnote {Present address:
Kamerlingh Onnes Laboratory, Leiden University, 2300 RA Leiden,
The Netherlands}} \affiliation{Institute of General Physics,
Russian Academy of Sciences, 119991 Moscow, Russia}

\begin{abstract}
Complex conductivity spectra of superconducting Nb are calculated
from the first principles in the frequency region around the
energy gap and compared to the experimental results. The row
experimental data obtained on thin films can be precisely
described by these calculations.\\

Submitted to Solid State Communications: 29 October 2003,
accepted: 12 November 2003\\

\end{abstract}

\maketitle

Observations of the superconducting energy gap in the far-infrared
spectra of conventional superconductors have played an important
role in establishing the Bardeen-Cooper-Schrieffer (BCS) theory of
superconductivity \cite{tinkham}, and provided an evidence for
importance of strong-coupling effects predicted by Eliashberg and
Holstein theories \cite{palmer}.

Despite numerous studies \cite{richards, anlage, klein,
marsiglio}, for Nb the influence of the strong coupling on optical
properties is not as clear as it is for Pb or Hg, for instance.
Most recent reports on experimental results of the surface
impedance in the microwave region are rather contradictory. Anlage
\textit{et al.} \cite{anlage} and Klein \textit{et al.}
\cite{klein} described their data basically by the BCS theory,
while Marsiglio \textit{et al.} \cite{marsiglio} were not able to
fit their results either by the simple BCS model or by
calculations according to the Eliashberg theory\ cite{eliash} with
the behavior of $\alpha^{2}F(\omega)$ evaluated from tunnelling
measurements \cite{arnold}.

Only a few experimental studies \cite{nuss, pronin} determine both
components of the complex conductivity,
$\sigma(\omega)=\sigma_{1}(\omega)+i\sigma_{2}(\omega)$, as a
function of frequency in the most interesting frequency region -
around the superconducting energy gap $2\Delta$, that allows to
make quantitative comparison with theoretical strong-coupling
calculations.

In one of such experiments \cite{pronin}, both, the real
$\sigma_{1}(\omega)$ and the imaginary $\sigma_{2}(\omega)$ parts
of the complex conductivity have been directly determined from the
transmission and the phase shift spectra of high-quality thin
niobium films on transparent substrates. The measurements have
been performed at several temperatures above and below $T_{c}$ in
the frequency range 5 to 30 cm$^{-1}$, precisely the region where
the superconducting gap lies. It has been shown that the overall
frequency dependence of the conductivity can be described using
the BCS formalism and assuming finite scattering effects. At the
lowest temperatures, however, some deviations from the BCS
predictions have been observed in the frequency behavior of the
complex conductivity. The deviations in the real part of
$\sigma(\omega)$ could not be explained by the strong-coupling
effects. In a subsequent comment \cite{halbritter}, an attempt has
been undertaken to explain the reported deviations by the granular
structure of the sputtered Nb films. These films have been
considered as a two-component system with an intrinsic BCS type
conductivity $\sigma_{BCS}$ and a grain-boundary conductivity
$\sigma_{bn}$ of Josephson type.

In this communication we show that the experimentally measured in
Ref. \cite{pronin} data for niobium, can be very accurately
described using the first-principles approach developed in Ref.
\cite{savrasov}. No Josephson junctions inside the sample are
needed to be considered.

This first-principles approach (for a review see \cite{maksimov})
allows to calculate the electron and the phonon spectra of metals
and the spectral functions of the electron-phonon interaction. As
a result, many related to the electron-phonon interaction
properties, such as the temperature-dependent electrical and
thermal resistivity, the plasma frequencies, the electron-phonon
coupling constants, \textit{etc.}, have been computed in Ref.
\cite{savrasov}.

For Nb, it has been shown that the electron-phonon contribution to
the dc resistivity $\rho$(T) is well described by the
first-principles calculated values of the plasma frequency
$\omega_{pl}$=9.47 eV and of the transport electron-phonon
coupling constant $\lambda_{tr}$=1.17 \cite{savrasov1}.

The resistivity $\rho$(T) of niobium can be written as:
\begin{equation}
\label{rho} \rho(T)=\frac{4\pi}{\omega_{pl}^{2}}\gamma(T),
\end{equation}
where $\gamma$(T) is the relaxation rate:
\begin{equation}
\label{gamma} \gamma(T)=\gamma_{imp}+\int^{\infty}_{0}
d\Omega\alpha_{tr}^{2}(\Omega)F(\Omega)\frac{\Omega/2\pi T}
{\sinh^{2}(\Omega/2\pi T)}.
\end{equation}
The first term here is the relaxation rate due to the temperature
independent impurity scattering. The second term describes the
temperature dependent contribution from the electron-phonon
interaction.

At considerably high temperatures (T$\ > 0.2\Theta_{D}$, here
$\Theta_{D}$ is the Debay temperature) Eq. \ref{gamma} can be
simplified to:
\begin{equation}
\label{gamma1} \gamma(T)=\gamma_{imp}+2\pi\lambda_{tr}T,
\end{equation}
where $\lambda_{tr}$ is the transport constant of the
electron-phonon coupling:
\begin{equation}
\label{lambda_tr} \lambda_{tr}=2\int^{\infty}_{0}
\alpha_{tr}^{2}(\Omega)F(\Omega)\frac{d\Omega}{\Omega}.
\end{equation}

\begin{figure}
\centering
\includegraphics[width=\columnwidth,clip]{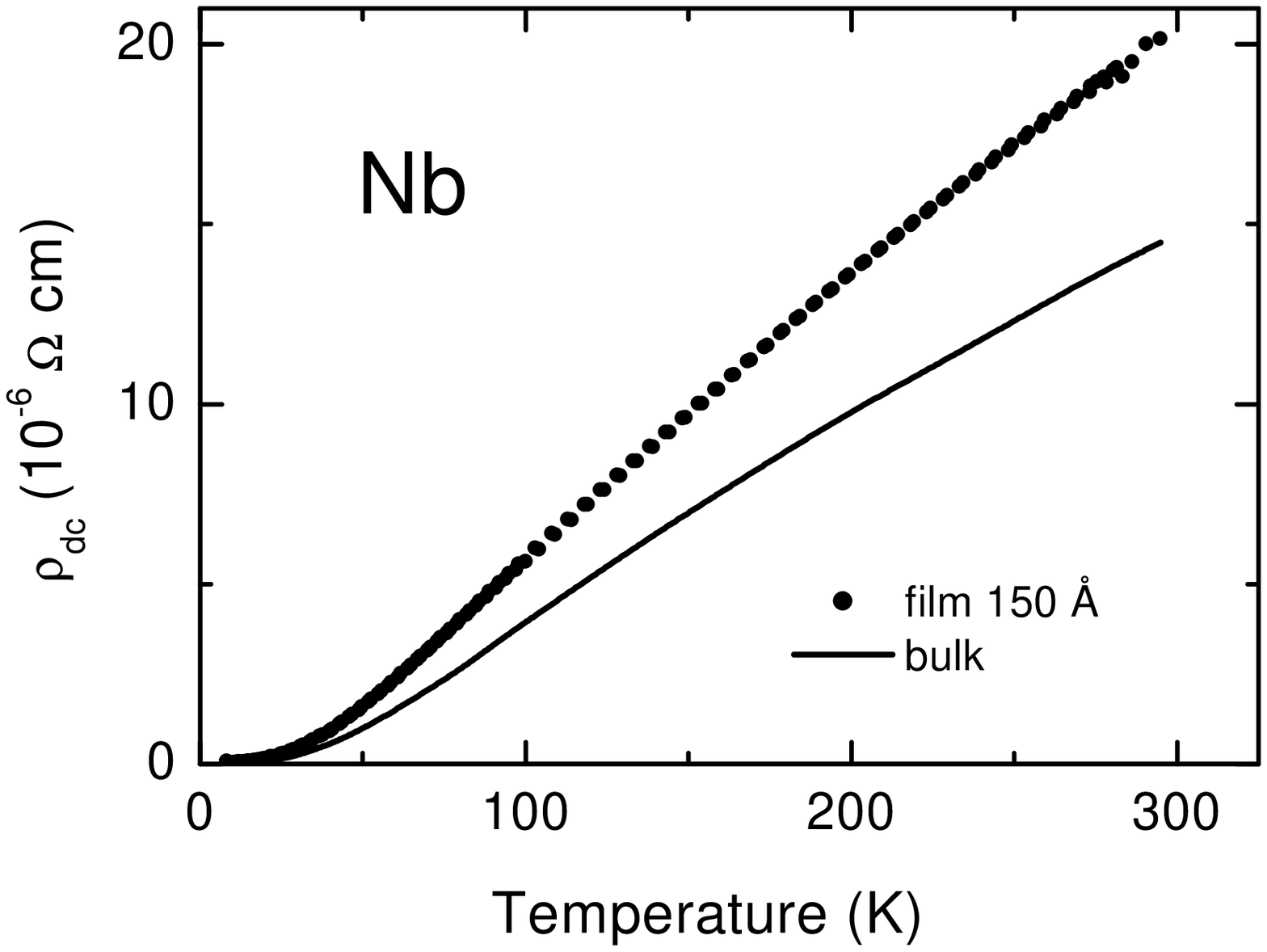}
\vspace{0.2cm} \caption{Temperature dependence of phonon part of
dc resistivity for thin-film \cite{pronin} and bulk \cite{bulk}
niobium.} \label{rho_dc}
\end{figure}

Fig. \ref{rho_dc} shows temperature dependent part of the dc
resistivity of the thin magnetron-sputtered Nb film \cite{film} in
comparison with the literature data for bulk Nb \cite{bulk}. The
residual impurity resistivity of the film is $\rho_{0} = 4.1
\times 10^{-6} \Omega$cm \cite{pronin}. As it is seen from the
figure, the experimental data obtained on a very thin (150 \AA)
film do not demonstrate significate difference from the bulk
results. The slope of the film curve is somewhat stepper than this
for the bulk curve. A little decrease of a plasma frequency (by no
more than 10\% of the bulk value) can describe this difference.

The critical temperature of the film, $T_{c}$ = 8.31 K
\cite{pronin}, is also somewhat below the bulk value (9.3 K),
although for films of this thickness, it is rather high
\cite{greene}. The influence of the film thickness on the
properties of thin films in the normal and superconducting states
is certainly beyond the scope of this work. For detail discussions
on this topic we refer to Ref. \cite{andreone} and references
therein. Here we just emphasize that the deviations of all the
parameters for the film measured in Ref. \cite{pronin} from the
bulk values are within $\pm10\%$, thus the investigated film can
pretty accurately represent the bulk properties.

From the theoretical point of view, these small differences
between the film and bulk properties can be understood, for
instance, in terms of a simple approach developed by Testardi and
Matteiss \cite{testardi}. This approach introduces smearing of the
first-principles calculated band structure of metals due to the
lifetime effects, which are certainly very essential in thin
films, and allows to calculate the density of states $N(0)$ and
the plasma frequency as function of the phonon ($\rho_{ph}$) and
the residual ($\rho_{0}$) resistivity. Although the
Testardi-Matteiss approach is oversimplified, it gives the right
direction of the change in the film properties compared to the
bulk.

For calculation of the frequency dependent conductivity of Nb, we
have used the formulas, obtained firstly by Nam \cite{nam}. Here
we present them in a shape convenient for numerical computations:

\begin{eqnarray}
\label{nam} \sigma(\omega)=\frac{\omega_{pl}^{2}}{4\pi i
\omega}\cdot\{
\int_{0}^{\omega/2}d\omega'[f(\omega')-f(\omega'-\omega)]\nonumber \\
\times\{\frac{1+n(\omega-\omega')
n(\omega')-a(\omega-\omega')a(\omega')}{\varepsilon(\omega-\omega')+\varepsilon(\omega')+
2i\gamma_{imp}} \nonumber \\
-\frac{1-n(\omega-\omega')
n^{*}(\omega')+a(\omega-\omega')a^{*}(\omega')}{\varepsilon(\omega-\omega')-
\varepsilon^{*}(\omega')+2i\gamma_{imp}}\} \nonumber \\
+\int_{0}^{\infty}d\omega'[f(\omega')-f(\omega'+\omega)] \nonumber \\
\times\{\frac{1-n(\omega+\omega')
n(\omega')-a(\omega+\omega')a(\omega')}{\varepsilon(\omega+\omega')+\varepsilon(\omega')+
2i\gamma_{imp}} \nonumber \\
-\frac{1+n(\omega+\omega')
n^{*}(\omega')+a(\omega+\omega')a^{*}(\omega')}{\varepsilon(\omega+\omega')-
\varepsilon^{*}(\omega')+2i\gamma_{imp}}\} \nonumber \\
-\int_{0}^{\infty}d\omega'\tanh(\frac{\omega+\omega'}{2T}) \nonumber \\
\times Re\frac{1-n(\omega+\omega')
n(\omega')-a(\omega+\omega')a(\omega')}{\varepsilon(\omega+\omega')+\varepsilon(\omega')+
2i\gamma_{imp}} \nonumber \\
-\int_{0}^{\omega/2}d\omega'\tanh(\frac{\omega-\omega'}{2T}) \nonumber \\
\times Re\frac{1-n(\omega-\omega')
n^{*}(\omega')+a(\omega-\omega')a^{*}(\omega')}{\varepsilon(\omega-\omega')-
\varepsilon^{*}(\omega')+2i\gamma_{imp}}\},
\end{eqnarray}
where $f$ is the Fermi distribution and
\begin{eqnarray}
\label{nam1}
\varepsilon(\omega)=Z(\omega)\sqrt{\omega^{2}-\Delta^{2}(\omega)}, \nonumber \\
n(\omega)=\frac{\omega}{\sqrt{\omega^{2}-\Delta^{2}(\omega)}}, \nonumber \\
a(\omega)=\frac{\Delta}{\sqrt{\omega^{2}-\Delta^{2}(\omega)}}.
\end{eqnarray}

The renormalization function $Z(\omega)$ and the energy gap
$\Delta(\omega)$ have been calculated from the Eliashberg equation
\cite{eliash}, which has a well-known form, and we do not
reproduce it here. In our calculations, we used the Eliashberg
spectral function $\alpha^{2}F(\omega)$ obtained from the
first-principles calculations in Ref. \cite{savrasov}.

\begin{figure}
\centering
\includegraphics[width=7cm,clip]{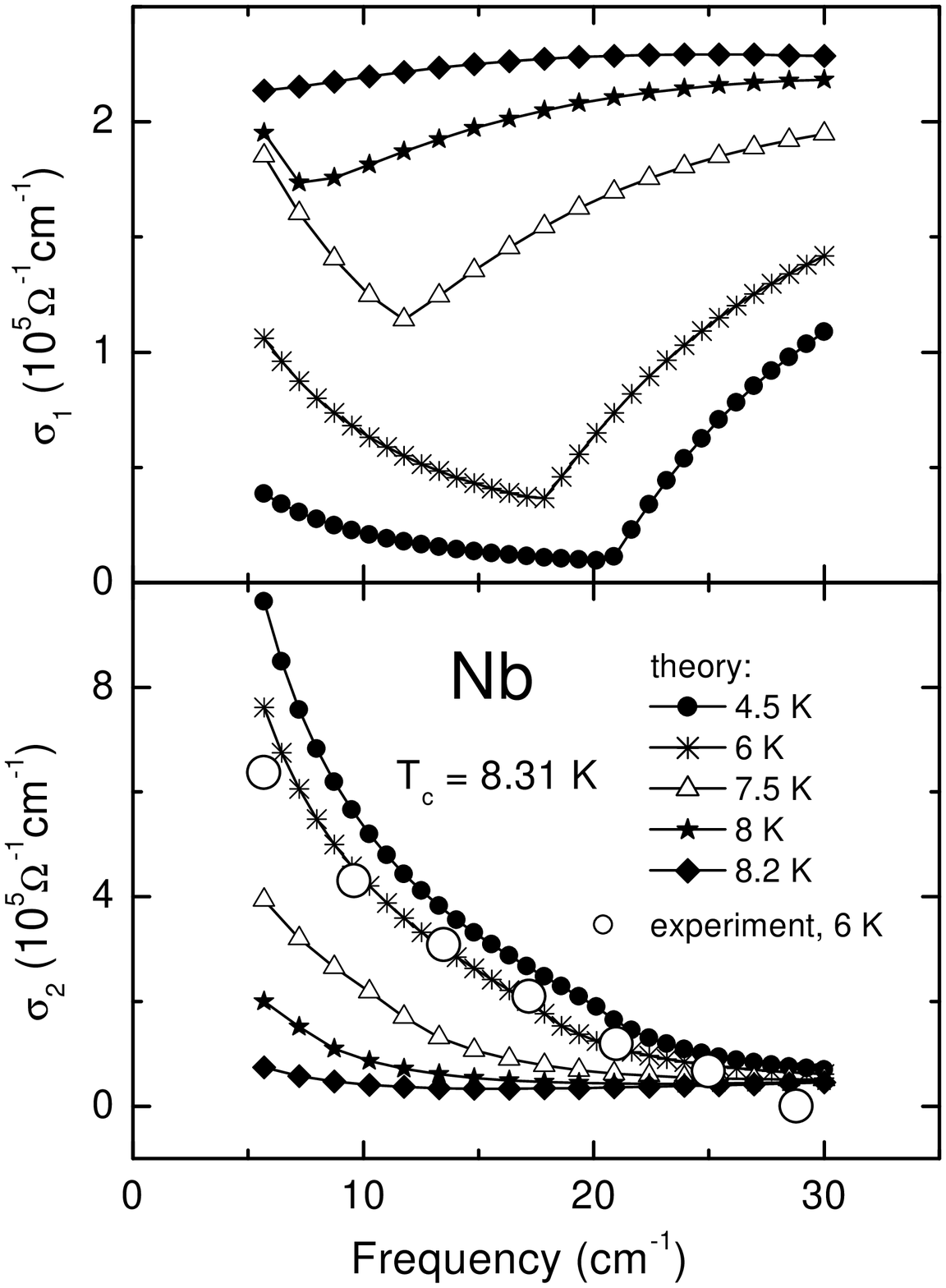}
\vspace{0.2cm} \caption{Frequency dependence of the real (upper
panel) and imaginary (bottom panel) parts of the complex
conductivity of Nb, as calculated from the first principles.
Compare to Fig. 4 of Ref. \cite{pronin}. As an example, we replot
the experimental $\sigma_{2}(\omega)$ data from this figure at T =
6 K.} \label{sigma12}
\end{figure}

In order to match to the experimentally measured critical
temperature of the film, we had to slightly increase the Coulomb
pseudopotential $\mu^{*}$ in the Eliashberg equation, in
comparison with the bulk value. As films are usually more
disordered than bulk samples, and as the disorder is known to
increase the value of $\mu^{*}$, this increasing seems to be well
grounded \cite{sadovskii}. Due to the reasons discussed above,
while computing the Eq. \ref{nam}, the plasma frequency has also
been slightly changed from its bulk value of 9.47 eV to 9.2 eV.

The real and imaginary parts of the complex conductivity, as
obtained from our calculations, are shown in Fig. \ref{sigma12}
for several temperatures below $T_{c}$. At each temperature, the
calculations have been made for 20 - 30 frequency points in the
interval from 5 to 30 cm$^{-1}$. As temperature goes down, the
energy gap clearly develops in the $\sigma_{1}(\omega)$ spectra,
$\sigma_{2}(\omega)$ demonstrating a characteristic
$1/\omega$-divergency at $\omega \rightarrow 0$.

\begin{figure}
\centering
\includegraphics[width=7cm,clip]{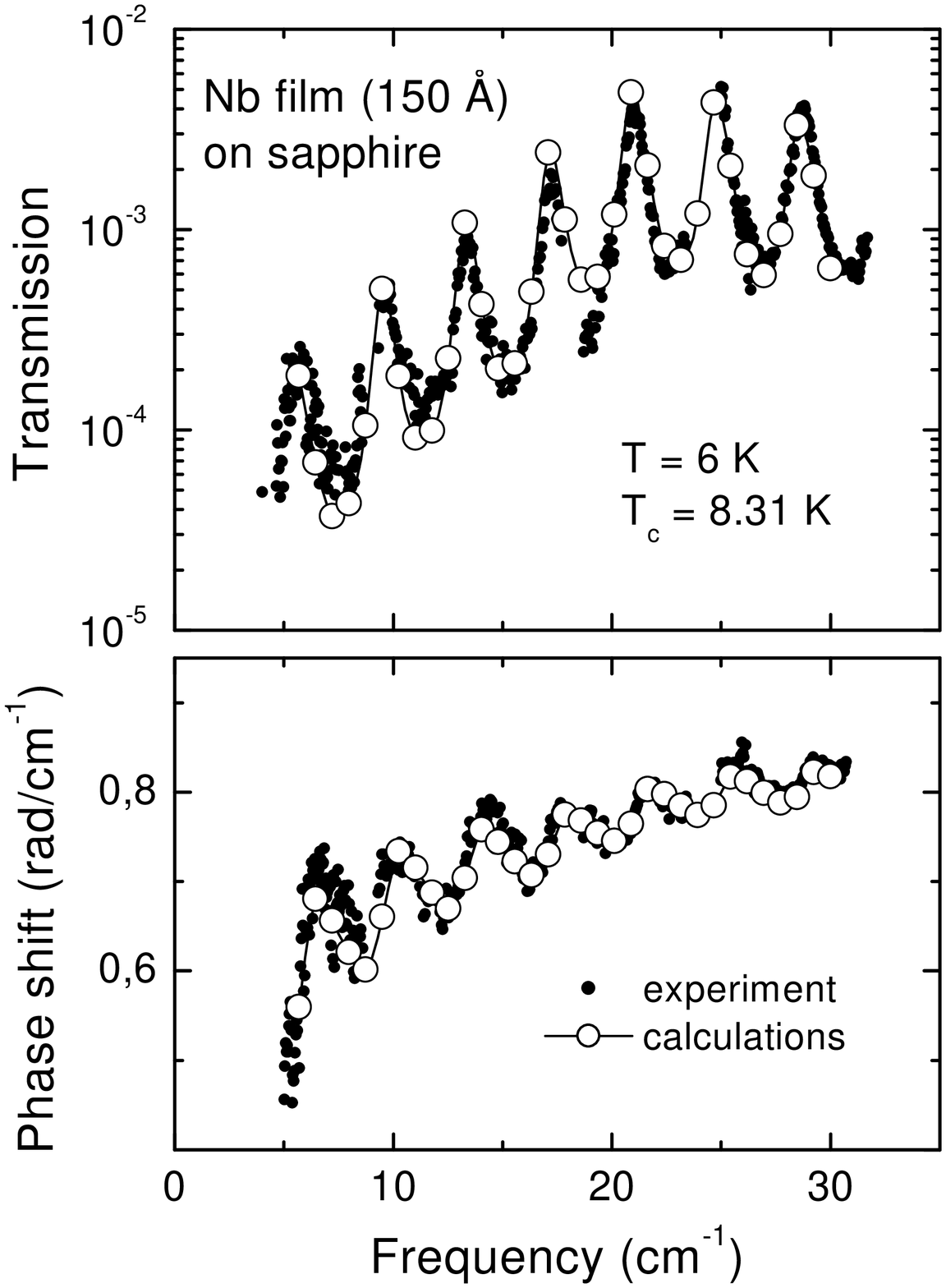}
\vspace{0.2cm} \caption{Examples of the transmission and phase
shift spectra of a 150-\AA-thick Nb film on sapphire substrate.
Open symbols - spectra, calculated from the first principles.
Solid symbols - experimental data from Ref. \cite{pronin}.}
\label{tr_pt}
\end{figure}

To compare our computations with the raw experimental data, we
calculate the transmission and the phase shift spectra of a Nb
film on sapphire substrate (as has been measured experimentally in
Ref. \cite{pronin}). For these calculations, the optical
parameters of the substrate and the thicknesses of the film and of
the substrate have been taken from the same reference. The
calculations were performed at all temperatures where the
experimental data were available. An example of these calculations
together with the experimental data are shown in Fig. \ref{tr_pt}
for $T = 6$ K. Clearly, the computed spectra beautifully describe
row experimental results.

In Fig. \ref{sigma12}, as large open symbols, we show $\sigma_{2}$
data, calculated in Ref. \cite{pronin} from the experimentally
measured transmission and phase shift spectra for $T = 6$ K. Our
first-principles calculations fit these experimental points very
well, while the weak-coupling calculations have been shown to
significantly deviate from the experimental values, especially at
the lowest frequencies (Fig. 5 of Ref. \cite{pronin}). Thus, the
effect of strong coupling is clearly seen in the
frequency-dependent conductivity in Nb.

We should note, that at the lowest temperatures and low
frequencies, the real part of the complex conductivity, calculated
in Ref. \cite{pronin} from the measured transmission and phase
shift spectra, does not coincide with our first-principles
calculations. We suggest this deviation is caused by the
difficulties in solving the inverse task - calculation of
$\sigma_{1}(\omega)$ and $\sigma_{2}(\omega)$ from transmission
and phase shift. At the lowest temperatures and at frequencies
below the gap, the imaginary part of conductivity is much large
than the real part. Consequently, the experimentally measured
dynamic response (transmission and phase shift in our case) is
mainly determined by $\sigma_{2}(\omega)$, while
$\sigma_{1}(\omega)$ has quite a minor influence on experimental
spectra. That causes large error bars in determination of
$\sigma_{1}(\omega)$ from experimental data.

Since the relaxation rate (which can be determined from Eqs.
\ref{rho} - \ref{gamma}) is much larger than the gap value, or, in
other words, the mean free path is smaller than the coherence
length, the film is in the dirty limit and the following
expression for the penetration depth is valid:
\begin{equation}
\label{lambda} \lambda^{-2}(T)=\sigma_{n}\cdot 2\pi
Im\int^{\infty}_{0}
\frac{\Delta^{2}(\omega)\tanh(\omega/2T)d\omega}{\omega^{2}-
\Delta^{2}(\omega)},
\end{equation}
where $\sigma_{n}$ is the normal state conductivity:
\begin{equation}
\label{sig_n} \sigma_{n}=\frac{\omega_{pl}^{2}}{4\pi\gamma_{imp}}.
\end{equation}
For $T = 4$K, that gives $\lambda = 880$ \AA, in a good agreement
with the experimental result, $\lambda^{exp} = 900$ \AA
\cite{pronin}.

To make better comparison with the experimental results of Ref.
\cite{pronin}, we now compute the London penetration depth. By the
London penetration depth we mean the penetration depth that would
be observed in our superconductor, if it were in the clean limit.
For this goal, we shall put $\gamma_{imp}$ = 0 in Eqs. \ref{nam},
what gives:
\begin{equation}
\label{lambda_l}
\lambda^{-2}_{L}=\frac{\omega_{pl}^{2}}{2\pi}\cdot
Re\int_{0}^{\infty}
\frac{\Delta^{2}(\omega)\tanh(\omega/2T)d\omega}{(\omega^{2}-
\Delta^{2}(\omega))^{3/2} \cdot ReZ(\omega)}
\end{equation}
For T = 4 K we obtain $\lambda_{L}$ = 354 \AA, again in a perfect
agreement with the experimental result, $\lambda_{L}^{exp}$ = 350
\AA \cite{pronin}.

In conclusion, we have demonstrated that the row experimental data
on the frequency dependent conductivity of a thin Nb film,
obtained in Ref. \cite{pronin}, can be accurately described by the
first-principles calculations made for bulk Nb with just a small
($\pm10\%$) varying of such parameters as the plasma frequency and
the Coulomb pseudopotential. These deviations are most likely
connected to the disordered nature of films, and can be understood
in terms of the a simple theoretical approach \cite{testardi}. No
Josephson junctions inside the sample are needed to be considered
to describe the experimental findings. For the first time, in our
knowledge, the first-principles calculations of the complex
conductivity of a superconductor, have been successfully compared
with the experimentally measured data in the frequency region
around the superconducting gap.

We are very grateful to many people for fruitful discussions,
especially to O. Dolgov. This work has been partially supported by
RFBR (grants Nos. 01-02-16719 and 03-02-16252), by the Program of
the Presidium of RAS "Macroscopic Quantum Mechanics", and by the
Program of the Physical Science Division of RAS "Strongly
correlated electrons". E.G.M. is grateful to Prof. J.J.M. Braat
for the kind hospitality during the visit to the TU Delft, where a
part of this work has been done.


\begin{thebibliography}{99}
\bibitem{film}  The absolute values for the film dc resistivity are shown
in Fig. 1 of Ref. \onlinecite{pronin}. The residual resistivity
ratio $\rho_{300K}/\rho_{10K}$ is equal to 6.5, demonstrating high
quality of the film.
\end{thebibliography}
\end{document}